\documentclass[twocolumn]{aastex631}

\usepackage{natbib}

\defcitealias{2018ApJ...868...17A}{ABW}
\defcitealias{2009ApJ...705..659A}{ABB}

\graphicspath{{./}{figures/}}

\received{\today}
\revised{\dots}
\accepted{\dots}

\shorttitle{A Robust Test of the Existence of PBHs in Galactic DM Halos}
\shortauthors{Abramowicz et al.}

\begin{document}

\title{ 
A Robust Test of the Existence of Primordial Black Holes in Galactic Dark Matter Halos 
}
       

\author[0000-0003-0067-5895]{Marek Abramowicz}
\affiliation{Research Center for Computational Physics and Data Processing; Institute of Physics, Silesian University in Opava, Czech Republic}
\affiliation{Nicolaus Copernicus Astronomical Centre, Polish Academy of Sciences, Bartycka 18, 00-716 Warsaw, Poland}
\affiliation{Department of Physics, G{\"o}teborg University, SE-412-96 G{\"o}teborg, Sweden}
\email{marek.abramowicz@physics.gu.se}

\author[0000-0002-4991-8213]{Micha{\l} Bejger}
\affiliation{INFN Sezione di Ferrara, Via Saragat 1, 44122 Ferrara, Italy}
\affiliation{Nicolaus Copernicus Astronomical Centre, Polish Academy of Sciences, Bartycka 18, 00-716 Warsaw, Poland}
\email{bejger@camk.edu.pl}

\author[0000-0001-5207-5619]{Andrzej Udalski}
\affiliation{Astronomical Observatory, University of Warsaw,
Al. Ujazdowskie 4, 00-478 Warszawa, Poland}
\email{udalski@astrouw.edu.pl}

\author[0000-0002-8635-4242]{Maciek Wielgus}
\affiliation{Max-Planck-Institut f\"ur Radioastronomie, Auf dem H\"ugel 69, D-53121 Bonn, Germany}
\email{maciek.wielgus@gmail.com}

\begin{abstract}
If very low mass primordial black holes (PBH) within the asteroid/moon-mass range indeed reside in galactic dark matter halos, they must {\it necessarily} collide with galactic neutron stars (NSs). These collisions must, {\it again necessarily}, form light black holes (LBHs) with masses of typical NSs, $M_{\rm LBH} \approx \,1-2\,M_{\odot}$. LBHs may be behind events already detected by ground-based gravitational-wave detectors (GW170817, GW190425, and others such as a mixed stellar black hole--neutron star mass event GW191219\_163120), and most recently by microlensing (OGLE-BLG-2011-0462). Although the status of these observations as containing LBHs is not confirmed, there is no question that gravitational-wave detectors and microlensing are in principle {\it and in practice} capable of detecting LBHs. 
We have calculated the creation rate of LBHs resulting from these light primordial black hole collisions with neutron stars. On this basis, we claim that if improved gravitational-wave detectors and microlensing statistics of the LBH events would indicate that the number of LBHs is significantly lower that what follows from the calculated creation rate, then this would be an {\it unambiguous proof} that there is no significant light PBH contribution to the galactic dark matter halos. Otherwise, if observed and calculated numbers of LBHs roughly agree, then the hypothesis of primordial black hole existence gets strong observational support, and in addition their collisions with neutron stars may be considered a natural creation channel for the LBHs, solving the problem of their origin, as it is known that they cannot be a product of standard stellar evolution. 
\end{abstract}

\keywords{black holes -- primordial black holes -- dark matter -- neutron stars}

\section{Introduction} 
\label{sec:intro}

A hypothesis that the galactic dark matter (DM) halos may be partially composed of black holes (BHs) of primordial origin (primordial black holes, PBHs; see e.g., \citealt{Khlopov2010}) has several fundamental aspects that have been studied by numerous authors, as described, e.g., in the recent reviews by \cite{2021arXiv211002821C,2021RPPh...84k6902C,2022arXiv220502153F}. In this Letter we report on one key issue that necessarily arises in the PBH context, namely, the formation of low-mass black holes (LBHs), that is, BHs with masses in the typical neutron star (NS) mass range, $M_{\rm LBH} \approx 1-2\,M_{\odot}$, resulting from collisions of light (asteroid/moon-mass, $10^{25}\,\mathrm{g} > M_{\rm PBH} > 10^{17}\,\mathrm{g}$) PBHs with NSs. We note here that the abundance of PBHs in this mass range is currently effectively unconstrained.

\cite{2018ApJ...868...17A} (hereafter \citetalias{2018ApJ...868...17A}) have demonstrated, using previous work by \cite{2009ApJ...705..659A} (hereafter  \citetalias{2009ApJ...705..659A}), that if light PBHs constitute a non-negligible fraction of galactic DM halos, their existence {\em must} have robust and inevitable consequences: they necessarily collide with galactic NSs, nest in their centers, accrete their dense matter interior, and eventually convert the NSs into LBHs, while releasing (possibly) observable electromagnetic (EM) signatures; \citetalias{2018ApJ...868...17A}, in their Equation 22, estimated the NS$\to$LBH conversion time to be inversely proportional to the initial PBH mass, assuming Bondi accretion approximation, which means that the accretion time may take days or years, depending on $M_{\rm PBH}$ (see also  \citealt{2021PhRvD.103j4009R} for results of numerical calculations). 

Various aspects pertaining to the fate of a NS colliding with an asteroid/moon-mass PBH was studied in the past. \cite{2017PhRvL.119f1101F} studied the process from the point of view of the r-process nucleosynthesis; specifically, an imploding rotating NS would spin up and shed a part of its mass, estimated to be between 0.1 and 0.5 $M_\odot$. \cite{2013PhRvD..87l3524C} considered captures of PBHs by NSs in the globular cluster cores. \cite{2014JCAP...06..026P} studied energy exchange in the close encounters of PBHs with NS, with possibly detectable gravitational-wave (GW) emission. \citet{2020PhRvD.102h3004G} studied in detail the process of PBH captures into NSs, providing a number of novel, potentially observable features in EM and GW domains, while \cite{2021PhRvD.104l3033K} revisited the \citetalias{2018ApJ...868...17A}'s idea that PBH-NS interaction may be connected with the fast radio burst emission. However, neither of these works put main focus on the fact that the very existence of LBHs would be an actual ''smoking gun'' evidence for the light PBHs, as the LBHs cannot be created in the standard stellar evolution. A~possibility of LBH formation in PBH collisions with low-metalicity stars was recently rediscovered by \cite{2022arXiv220513003O}. LBHs were also directly proposed in the context of recent observations, e.g., \cite{2021JCAP...10..019T} considered an alternative scenario for the GW170817 event \citep{PhysRevLett.119.161101}, in which one of the components was assumed to be a LBH; such alternative scenario is currently not ruled out by the state-of-art understanding of the multi-messenger physics of the merger (see e.g., \citealt{PhysRevD.100.043011}). 

 Inspired by recent discoveries of low-mass compact objects in GWs (\citealt{2021arXiv211103606T}, lighter component of GW191219\_163120 with a mass of $1.17^{+0.07}_{-0.06}\ M_\odot$) and microlensing observations (\citealt{2022arXiv220113296S,2022arXiv220201903L}, lower limit on the mass of the OGLE-2011-BLG-0462 lens estimated to $1.6\ M_\odot$), in Section~\ref{sec:brief_summary} we recast a summary of the \citetalias{2018ApJ...868...17A} arguments, leading to the calculation of the number of the LBH events in Section~\ref{sec:new_results}, with a discussion of the potential for breakthrough discoveries by future GW and microlensing observations contained in Sections~\ref{sec:gw} and \ref{sec:ml}, respectively, and conclusions gathered in Section~\ref{sec:conclusions}. 

\section{Collisions of hypothetical light PBH with galactic NS} 
\label{sec:brief_summary}  
 
If PBHs exist, they {\em must} collide with galactic objects. Single collisions are virtually unobservable, as \citetalias{2009ApJ...705..659A} demonstrated. However, a collision of PBH with an NS necessarily initiates observable consequences: a PBH hits NS nearly head on, with impact velocity nearly equal the escape velocity, goes through the NS, losing a small fraction of its kinetic orbital energy through dynamical friction \citep{1971ApJ...165....1R,1999ApJ...513..252O} and emerges out on the opposite side of the NS with velocity smaller than the escape velocity. This passage bounds a PBH to an NS gravitationally. The PBH therefore turns back and hits the NS again, losing orbital energy until it settles at the center of the NS and starts to accrete the NS matter at an increasing rate as its own mass grows. This eventually induces a conversion of the entire NS and formation of an LBH. It is important to stress that this sequence of events -- multiple hits, capture in the centre, Bondi accretion, dynamical collapse to an LBH -- is an {\it unavoidable} consequence of fundamental physics, as shown in the \citetalias{2018ApJ...868...17A} paper.

How often do LBHs form? \citetalias{2018ApJ...868...17A} constructed a special purpose population synthesis method to answer this question. The method uses a simplified model of a galaxy consisting of a bulge and a disk, reproducing the density profile required to explain the typically observed galactic velocity profile. The details of the model together with the parameters used are summarized in \citetalias{2018ApJ...868...17A}'s Section 2.2.

Matter content of a model galaxy presented in \citetalias{2018ApJ...868...17A} consists of five species: $i$$=$$1$ - primordial black holes (PBH), $i$$=$$2$ -  neutron stars (NS), $i$$=$$3$ - standard stellar-mass black holes (BH), $i$$=$$4$: - low mass black holes (LBH) resulting from the PBH--NS collisions, and $i$$=$$5$ - stars, interstellar gas and dust. The total local mass density does not change during the galaxy evolution. However, the volume number densities of the species $n_{(i)}(r)$ do change, as they interact, for example as NS and PBH collide and form together LBH. The evolution of the species $i=1$--$4$ is described by  
\begin{equation}
\frac{\partial n_{(i)}}{\partial t} = \sum_{k=1}^4 n_{(i)} n_{(k)}\,C^{(i)(k)} + K_{(i)},\quad k,i=1,2,3,4.
\label{eq-population-synthesis}
\end{equation}
\noindent Here, the collision coefficients $C^{(i)(k)}$ express the condition for a direct capture of species $k$ by species $i$, and follow from standard argument in classical mechanics based on energy and momentum conservation (it is recalled and explained in the present context by \citetalias{2009ApJ...705..659A}),
\begin{equation}
n_{(i)} n_{(k)}\,C^{(i)(k)} = n_{(i)} n_{(k)} \frac{V_{(i)}^2}{V(r)} R^2_{(i)}.
\label{eq-collision-coefficient}
\end{equation}  
\noindent Here $R_{(i)}$ is radius of the $i$ species and $V_{(i)}$ the escape velocity from it (it equals to the speed of light $c$ for the BH species). For $i=2,3$ the creation coefficients 
reflect the galactic supernova rate of 0.01\,yr$^{-1}$. We envisage a range of LBH masses $M_{\rm LBH}\approx 1-2\,M_{\odot}$, the details depending on the NS mass function and physics of the final NS$\to$LBH collapse. The LBH mass function will likely follow the NS mass function shifted toward smaller values, as some of the NS mass may be lost during the conversion \citep{2017PhRvL.119f1101F}. For practical purposes and simplicity in numerical calculations we represent the NS/LBH mass distributions with a single characteristic value  $M_{\rm NS} = M_{\rm LBH} = 1.5\,M_\odot$.


\section{Calculating the number of LBH events} 
\label{sec:new_results} 

\begin{figure}[h!]
\begin{center}
{\includegraphics[width=1.00\linewidth]{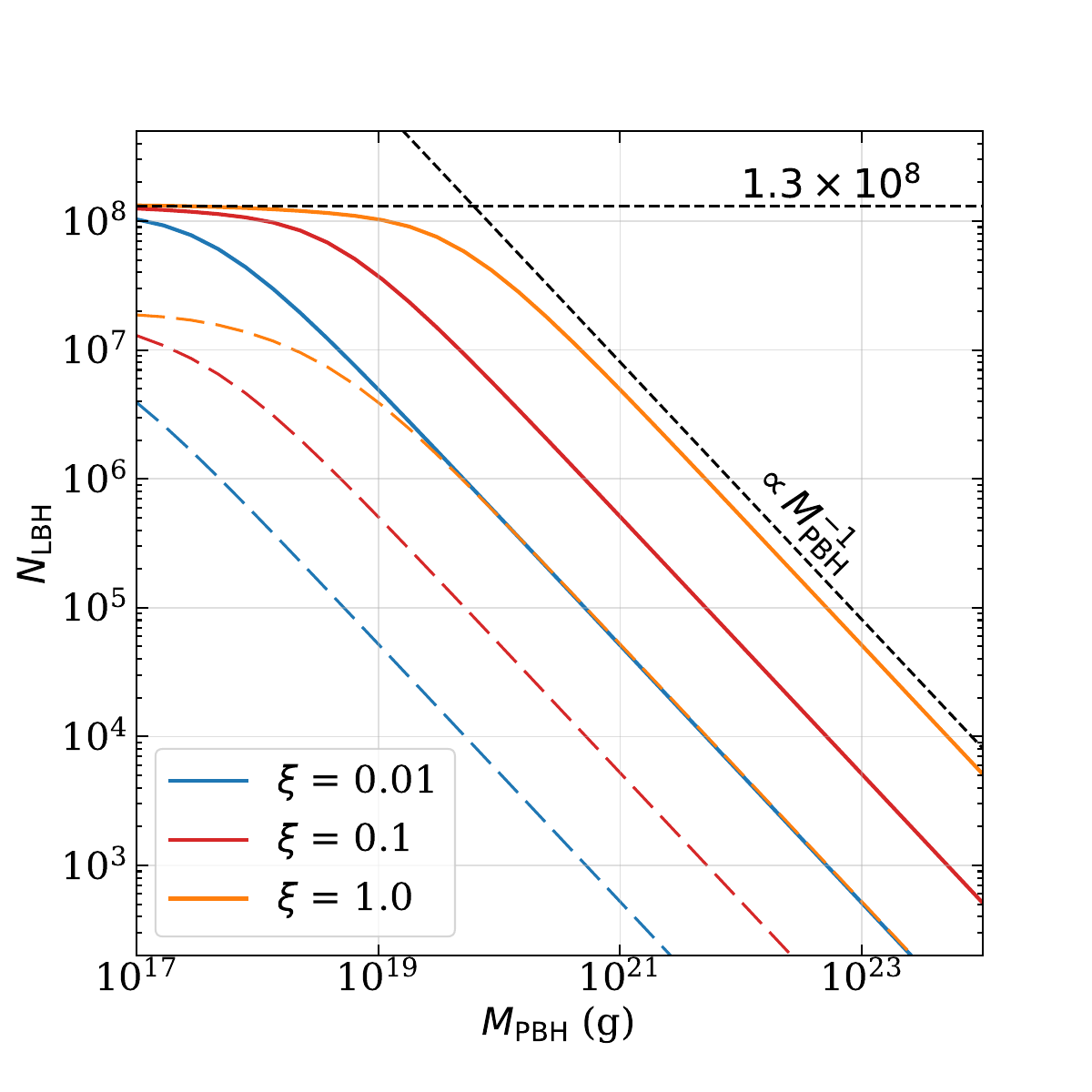}}
\caption{\small Absolute number of LBHs in a model galaxy of 13 Gyr age as a function of $M_{\rm PBH}$ and $\xi$. Majority of LBHs are created in the galactic bulge. The continuous lines correspond to the total integrated number of LBHs in the galaxy, while the dashed lines represent the LBHs created in the galactic disk. 
If one fixes $\xi$ and assumes $M_{\rm PBH}$ to be $n$ times smaller, then there is necessarily $n$ times more PBHs flying around, so it is $n$ times easier for a PBH to collide with an NS. Thus, there is about $n$ times more LBHs formed. Therefore, in the regime $N_{\rm NS} \gg N_{\rm LBH}$ we have the $N_{\rm LBH} \propto M_{\rm PBH}^{-1}$ scaling.
} 
\label{fig-absolute_numbers}
\end{center}
\end{figure}

The \citetalias{2018ApJ...868...17A} population synthesis method, which we employ here, is based on solving Equation~\ref{eq-population-synthesis}, and starts with the initial galaxy state with no NS, BH or LBH. The PBH represent a fraction $\xi$ of the total gravitating mass in the galaxy, assumed to be $\sim 7 \times 10^{10} M_{\rm \odot}$ in the model galaxy. The total number of LBHs derived from the \citetalias{2018ApJ...868...17A}'s model, as a function of characteristic PBH mass $M_{\rm PBH}$ and mass ratio $\xi$, is shown in Figure~\ref{fig-absolute_numbers}. For low $M_{\rm PBH}$ the curves approach the limit of total number of created NSs, for heavier PBHs there is an inverse dependence of $N_{\rm LBH}$ on $M_{\rm PBH}$.

\begin{figure*}[t]
\begin{center}
{\includegraphics[width=\linewidth]{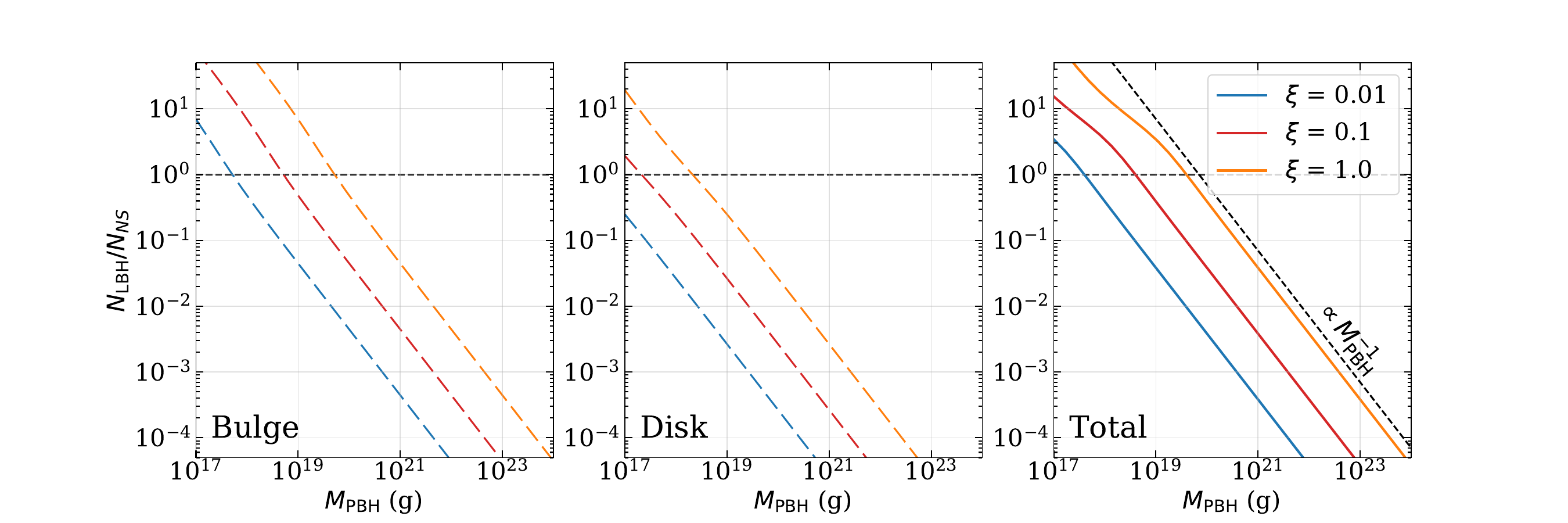}}
\caption{\small Ratio of the density of LBHs to NSs in a model galaxy of 13 Gyr age, as a function of $M_{\rm PBH}$, the characteristic mass of PBHs, for 3 different fractions of PBH in a galactic mass composition $\xi$. This ratio represents the expected ratio of occurrence of collisions of stellar-mass BHs with LBHs and stellar-mass BHs with NSs. The three panels correspond to the galactic disk, bulge, and the total galaxy. For large $\xi$ or light $M_{\rm PBH}$ majority of NSs may become LBHs.}
\label{fig-results-ABW-simulations}
\end{center}
\end{figure*}

A number that can be more practical from the observational point of view than the absolute number of LBHs is the ratio of the number of LBHs to the number of NSs. The results of these calculations are shown in Figure~\ref{fig-results-ABW-simulations}. The ratio of the number of LBHs to NSs is approximately inversely proportional to $M_{\rm PBH}$ for a fixed galactic mass PBH fraction $\xi$, with the model's simplistic assumption that the all PBHs have one characteristic mass $M_{\rm PBH}$. This example demonstrates that in principle $N_{\rm LBH}/N_{\rm NS}$ contains information on the $M_{\rm PBH}$, and therefore can be used to specify the $M_{\rm PBH}$ mass function, e.g., if all PBHs have small masses, $M_{\rm PBH}\approx 10^{17}\ \mathrm{g}$, then the majority of NSs should be converted to LBHs, as $N_{\rm LBH}/N_{\rm NS} > 1$. Any observational constraint on the $N_{\rm LBH}/N_{\rm NS}$ ratio will place limits on masses and abundance of PBHs.

From existing GW and microlensing constraints, NS-mass PBHs do not account for the majority of DM content, but may still be responsible for producing detectable events, e.g., in GWs \citep{2017PhRvD..96l3523A,2020PhRvD.101d3015V,2022PhRvD.105h3526F}. In the following Sections we mainly discuss ways to distinguish the NS-mass PBHs from {\em bona fide} NSs by means of the GW and microlensing observations, but also want to signal two thought-provoking subtleties. The first one concerns a distinction between LBHs created from NSs in a process of asteroid-mass PBH collision from genuinely primordial BHs. Distinguishing characteristics of these distinct classes may be the value of their spins, as discussed briefly in Section~\ref{sec:gw}. The second aspect is related to distinguishing LBHs produced by asteroid-mass PBHs from LBHs created by other possible DM candidates \citep{2018PhRvD..97e5016B,2021PhRvL.126g1101T,2022PhRvD.105l3030G}, or in other scenarios, e.g. in collapsing supramassive NSs \citep{Falcke2014}. One may expect different conversion processes, resulting in different final features, such as objects' masses, spins and their EM emission (see e.g., \citealt{2016arXiv160307830A}), which may in turn be related to possible existence of exotic objects such as gravastars, boson stars, or naked singularities (see \citealt{Cardoso_2019} for a review). A detailed discussion of these possibilities is beyond the scope of the present Letter, but any extraordinary observation of this sort would imply a signature of exciting new physics.

\subsection{Gravitational-wave constraints} 
\label{sec:gw} 

Detections of compact binary inspiral and merger events, including GWs from sources composed of NS-mass components - most notably GW170817, the only one with confirmed EM counterpart, but also GW190425 and ''mixed'' BH-NS-mass binaries GW200115\_042309, GW191219\_163120, GW190917\_114630 (see the GWTC-3 catalogue, \citealt{2021arXiv211103606T}) -- by the LIGO-Virgo-KAGRA (LVK) Collaboration, has revived a discussion on the primordial origin of some of the components, due to overall low spin distribution of the observed population and event rates consistent with PBHs \citep{Clesse:2020ghq}. According to models \citep{Luca_2019,Mirbabayi_2020} sufficiently heavy PBHs (masses larger than ${\sim}10\,M_\odot$), may acquire substantial spins in process of accretion prior to the reionization epoch \citep{Luca_2020a,Luca_2020b,2022PhRvD.105f3510F}, but lighter PBHs (specifically, NS-mass PBHs) should in general remain non-rotating. Additionally, existence of very light PBHs in specific mass ranges is difficult to exclude observationally, see e.g., \cite{2019JCAP...08..031M,2021arXiv211002821C,2021RPPh...84k6902C,2022arXiv220502153F}. As estimated in \cite{clesse2018seven}, already relatively massive dark compact objects (between $10^{-6}$ and $10^{-5}~M_\odot$) could constitute a fraction of DM of order 1\%, whereas asteroid/moon-mass PBHs (light PBHs) abundance is effectively unconstrained 
\citep{2019JCAP...08..031M,2021arXiv211002821C,2021RPPh...84k6902C,2022arXiv220502153F}. 

The only EM-bright GW detection so far, the GW170817 event, does not exclude the possibility of one component being a LBH (see e.g., \citealt{PhysRevD.100.043011}). Using the merger rate obtained by the LIGO and Virgo collaborations, based on these recent observations, \citetalias{2018ApJ...868...17A} approximated (see their Section 3.5) the fraction of events containing a LBH to be of the order of a percent of the whole binary merger population containing NS-mass objects.  

Clearly, a high signal-to-noise GW measurement which could prove the existence of LBHs would be of a binary LBH inspiral and merger, in which component masses $M_1$ and $M_2$ lying well within the NS mass range are measured; if LBHs are created in PBH-induced implosion event, their masses should be systematically lower than the progenitor NS masses due to mass-shedding spin up \citep{2017PhRvL.119f1101F}; consequently, their spins should be high, as estimated by \citetalias{2018ApJ...868...17A} (their Section 3.5). Just before the merger a binary LBH system would reach frequencies excluded by NS properties (compactness $M/R$, with $R$ denoting the NS radius, and consequently the equation of state), as the maximum final GW frequency can be estimated as $f^c_{GW} \approx c^3/(2\sqrt{2} G\pi)/(M_1 + M_2)$ for the case of two BHs \citep{2017AnP...52900209A}. If both components are LBHs, the best-matched waveform would indicate no measurable mass-weighted average tidal deformability (effective tidal deformability of the binary) $\tilde{\Lambda} = (16/13)\left((M_1 + 12M_2)M_1^4 \Lambda_1 + (M_2 + 12M_1)M_2^4 \Lambda_2\right)/ \\ \left(M_1 + M_2\right)^5$, with $\Lambda_i = k(2/3)\left(R_i/M_i\right)^5$ denoting individual components' tidal deformabilities, with $k$ denoting a functional of the equation of state. This effect should be especially evident for component masses near the low end of the NS mass spectrum (i.e. for low compactness $M/R$, high tidal deformability objects). For both LBHs being spacetime curvature objects, no detectable EM emission during and after the merger is expected. 

If at least one component is a LBH, at least a sub-population of them should possess a relatively high spin, which in turn should have an detectable influence on the detected waveform, i.e., be measurable by the effective binary spin $\chi_{\rm eff} = \left(M_1\vec{\chi_1} + M_2\vec{\chi_2}\right)\hat{L}/\left(M_1 + M_2\right)$, with $\chi_i$ individual spins, and $\hat{L}$ is a direction of the system's orbital momentum \citep{2021arXiv211103606T}. If a GW signal was emitted by a LBH-NS system, the resulting measurement of the tidal deformability $\tilde{\Lambda}$ would place a limit on the NS equation of state, as demonstrated in \cite{2018A&A...620A..69H}, since one of the components' $\Lambda\equiv0$, potentially creating a tension with already known features of the equation of state of dense matter. Alternatively, a tidal disruption event or an EM-precursor before the merger would be observed, with details depending on component masses and the equation of state \citep{2021arXiv211103686N}. For a mixed NS-LBH system, a merger and post-merger evolution, especially from the EM point of view, e.g., the kilonova emission \citep{1998ApJ...507L..59L} should be potentially differentiable from a binary NS case \citep{2019LRR....23....1M}; from the GW point of view, no long-lasting post-merger GW signal is expected, but rather a prompt collapse to a final BH (the only nearby GW merger so far, GW170817, was too far for the detectors' sensitivity at that time to give conclusive evidence, \citealt{GW170817-postmerger}). 

Planned increase in the GW strain sensitivity by a factor ${\approx}1.5$ of the Advanced LIGO \citep{LIGOScientific:2014pky} and Advanced Virgo \citep{VIRGO:2014yos} detectors for the O4 observing campaign (with a fourth detector, KAGRA \citealt{2021PTEP.2021eA101A}, joining the global network) scheduled to start in March 2023\footnote{15 June 2022 update of the LVK O4 observing run plans: {\tt https://observing.docs.ligo.org/plan/}.} promises few times larger sensitive volume and hundreds of new transient GW detections, among them at least a few NS-mass inspiral events \citep{KAGRA:2013rdx}. In addition, searches for hypothetical sub-solar mass inspirals, which may contain LBHs \citep{2021arXiv210912197T} and general searches for continuous GWs (e.g. \citealt{Miller:2020kmv,Miller:2021knj,2022arXiv220100697T}) may provide additional evidence or upper limits for PBHs. Farther future, with 3rd generation detectors, the Einstein Telescope (ET, \citealt{Punturo_2010}) and the Cosmic Explorer (CE, \citealt{2019BAAS...51g..35R}) will {\em guarantee hundreds of thousands} of NS-mass binary inspiral detections per year \citep{2020JCAP...03..050M}. Additionally, for a case of a solitary NS being converted into a LBH, a number of potentially observable GW and accompanying EM features was recently studied in \cite{2020PhRvD.102h3004G}.

In summary, future GW measurements of coalescing binaries with NS-mass components will contain information pertinent to the inner structure of these objects. Systematic evidence for negligible tidal deformabilities, large spins, as well as EM counterparts inconsistent with the standard NS evolution (solitary or in binary systems) would serve as indication for the existence of LBHs, and therefore, indirectly, of PBHs. 

\subsection{Microlensing and X-ray constrains} 
\label{sec:ml} 

Gravitational microlensing \citep{1986ApJ...304....1P} can be a useful tool for
constraining the population of LBHs. This phenomenon is a result of
bending of the light rays of a distant source passing near a lensing
object.  It depends on the mass of lensing body and is independent of
its luminosity. Thus, microlensing provides an unique opportunity to
determine a mass of a single lens, even if it is non-luminous and
dark. Indeed, the recent announcements of the discovery of dark black
hole mass lens in the gravitational microlensing event
OGLE-2011-BLG-0462 \citep{2022arXiv220113296S,2022arXiv220201903L} proves that the method is sound and reliable.

In practice a direct mass determination {\it via} microlensing is
complicated because of a degeneration of the lens mass, geometry of
the event and its kinematics. To lift this degeneracy three
observables must be measured to derive the lens physical parameters:
the Einstein ring crossing time, $t_E$, microlensing parallax,
$\pi_E$, and angular size of the Einstein ring, $\Theta_E$: {e.g.}, 
$M=\Theta_E/\kappa/\pi_E$ where $\kappa$ is a constant \citep{2000ApJ...542..785G}. 
Only in the very rare cases all three observables can be
derived directly from photometry of a single lens microlensing. $t_E$
is routinely measured for every microlensing event, $\pi_E$ can be
generally derived for long-lasting events ($t_E>50$~days) from a tiny
deviation of the observed light curve shape from theoretical one due
to the orbital motion of the Earth around the Sun, $\Theta_E$ -- only
for high magnification events when the finite source size effects can
be measured.  Fortunately, the duration of the microlensing events
depends on the mass of the lens and in the case of LBH mass range
predicted events should last several tens or hundreds days. Thus, in
the majority of cases of LBH mass microlensing photometry should
provide $t_E$ and $\pi_E$.

On the other hand the Einstein ring, $\Theta_E$, can be measured from an additional
effect caused by microlensing – the astrometric microlensing, i.e., the shift of the position of the source star image centroid
during the phenomenon. This requires very precise astrometry as the
astrometric effects are very small, of the order of a milliarcsec in
the case of LBH mass. Thus, the space astrometry (HST, Gaia JWST or
future astrometric missions) is required or, for very bright sources,
ground based interferometry. OGLE-2011-BLG-0462 was a best example of
such a synergy where the ground based photometry, mostly from the OGLE
survey \citep{2015AcA....65....1U} was used supplemented with the HST
astrometric monitoring of the event. This successful detection proves
that the detection of LBHs is in principle possible with the
gravitational microlensing method. However, one should remember that
the microlensing method can detect dark objects of the LBH masses but
it cannot in principle distinguish if the lens is an LBH or
NS. Nevertheless, some additional follow-up or all-sky survey
observations, e.g., the level of X-ray flux, may allow separating
LBH from NS, similarly as in the case of OGLE-2011-BLG-0462 \citep{2022arXiv220607480M}. Thus, large-scale microlensing hunt for
LBHs/NS combining photometry and astrometry should be in position to
shed some light on the putative population of LBHs.

Microlensing provides statistical constraints on the fraction of the
dark objects in the Galactic DM halo as a function of object mass. The early limits from the first generation microlensing surveys conducted in the direction of the LMC yielded an upper limit for the
fraction of the LBH mass range objects in the Galactic halo dark
matter at about 10\% based on OGLE-II and OGLE-III observations \citep{10.1111/j.1365-2966.2011.19243.x}. 
Much more extensive, unpublished yet,
analysis based on about 20 years long combined OGLE-III and OGLE-IV
dataset indicates, however, even more stringent limit -- of about
0.5\% (P. Mr{\'o}z – private communication). In the opposite direction
-- toward dense stellar regions of the Galactic center -- additional
contribution from disk dark stellar remnants dominates. It is
estimated that about 3\% of all observed microlensing events in this
direction can be caused by classical NSs and 1\% by classical BHs \citep{2000ApJ...535..928G}. 
Thus, potential microlensing detection of the LBH
mass range objects seems to be feasible. What part of them could be
real LBHs and what -- regular NSs -- remains an open question to be
answered when a significant sample of such detections is collected.


\section{Conclusions} 
\label{sec:conclusions} 

Current developments of observational multi-messenger astrophysics (GWs, microlensing, X-ray observations) present a potentially fundamental breakthrough opportunity to scrutinize the hypothesis of light PBHs contributing to the gravitational potentials of galactic DM halos, by indirectly proving their existence by means of the existence of LBHs. 

A key point of our argument is that the existence of PBHs in galactic DM halos necessarily implies formation of LBHs, i.e., BHs with NS mass, as shown by \citetalias{2009ApJ...705..659A} and \citetalias{2018ApJ...868...17A}. What follows, inspirals and mergers of LBHs with standard compact objects (stellar-mass BHs and NSs) as well as mergers of binary LBHs would produce transient GWs detectable by the current ground-based detectors (LVK) within a Gpc range (with a greater reach in the future, as the sensitivity of the detectors will increase). On the other hand, Galactic microlensing events involving potential LBHs as gravitational lenses are detectable in practice by optical observatories, and may be inspected by X-ray follow-up observations to reconfirm the BH nature of the lens. 

From sufficiently complete statistics of these events -- which seems to be a matter of years rather than decades -- one will definitely build up evidence to decide for or against these two following, mutually-exclusive possibilities: if the observed number of the LBH events would turn out to be inconsistent with the number calculated on the basis of the population synthesis presented in this paper, then the robust conclusion would be that there are no asteroid/moon-mass PBHs in the galactic DM halos. If, however, the observed number reasonably agrees with the calculated one, this would provide a very strong supporting evidence for the existence of PBHs.


\begin{acknowledgments}
{\em Acknowledgments.} AU thanks dr Przemek Mr{\'o}z for a fruitful discussion. MW and MB thank Alexandra Elbakyan for her contributions to the open science initiative. This work was partially supported by the Polish National Science Center grant no. 2016/22/E/ST9/00037. M.A. acknowledges the Czech Science Foundation grant No. GX21-06825X. The Authors are grateful to the anonymous Referee for their valuable comments.
\end{acknowledgments}

\bibliography{references}

\end{document}